\documentstyle[12pt,dina4p,epsfig]{article}
\title{
  \vskip-2cm
  {\baselineskip16pt
    \centerline{\normalsize \tt DESY 97-002 \hfill ISSN 0418-9833}
    \centerline{\normalsize \tt hep-ph/9701247 \hfill}
    \centerline{\normalsize \tt January 1997 \hfill}
  }
  \vskip2cm
  {\bf 
    Jet Shapes in $ep$ and $p\overline{p}$ Collisions \\
    in NLO QCD
  }
  \author{
    {M.\ Klasen$^a$, G.\ Kramer$^b$} \\[5mm]
    {$^a$ Deutsches Elektronen-Synchrotron DESY, Notkestra\ss{}e 85,}\\
    {     D-22607 Hamburg, Germany, e-mail: {\tt klasen@mail.desy.de}} \\[5mm]
    {$^b$ II. Institut f\"ur Theoretische Physik\thanks
          {Supported by Bundesministerium f\"ur Bildung und Wissenschaft,
          Forschung und Technologie, Bonn, Germany under Contract 05\,7HH92P(0)
          and EU Program "Human Capital and Mobility" through Network
          "Physics at High Energy Colliders" under Contract
          CHRX-CT93-0357 (DG12 COMA)} ,
          Universit\"at Hamburg,}\\
    {     Luruper Chaussee 149, D-22761 Hamburg, Germany,}\\
    {     e-mail: {\tt kramer@desy.de}}
    }
  \date{}
}
\begin{document}
\maketitle
\vspace{3cm}
\begin{abstract}
\thispagestyle{empty}
We have calculated jet shapes in low $Q^2$ $ep$ and $p\overline{p}$
collisions in perturbation theory at order $\alpha_s^3$
for the hard parton-parton processes. For the $\gamma p$
process resolved and direct contributions are superimposed. The
dependence of the jet shapes on transverse energy, rapidity, and
inner cone extension is studied. The numerical results of the
calculation are compared with recent data from ZEUS at HERA and
from CDF and D0 at the TEVATRON. Good agreement is achieved if the
problem of merging overlapping jets is taken into account by
varying the parameter $R_{\rm sep}$ as a function of transverse energy
and rapidity.
\end{abstract}
\newpage
\setcounter{page}{1}
\section{Introduction}
A large fraction of the final state in hadron-hadron,
electron-hadron (both high $Q^2$ and low $Q^2$, the latter also being
referred to as photon-hadron), and electron-positron
collisions consists of high energy jets. These jets have an
extended structure which can be studied experimentally and
theoretically. A possible measure of this structure is the jet
shape or jet profile, which depends on variables like transverse
energy and rapidity of the jets, the jet algorithm, and the
extension of jets.
The jet shape measure is the function $\rho(r,R,E_T,\eta)$ where $E_T$
is the transverse energy of the jet and $\eta$ is its rapidity with $\phi$,
its azimuthal angle, integrated out. This function $\rho$
measures the fractional $E_T$
profile, i.e.~given a jet sample with transverse energy $E_T$
defined with a cone radius $R$, $\rho(r,R,E_T,\eta)$
is the average fraction of the jet's  transverse energy that
lies inside an inner cone with radius $r < R$. \\

This jet shape $\rho$ for the special value $r=0.2$ has been measured for the
first time in $p\overline{p}$ collisions at $\sqrt{s}$
= 630 GeV by the UA1 collaboration \cite{X20} at the $Sp\overline{p}S$
collider. More detailed measurements of $\rho$ in $p\overline{p}$ collisions
at $\sqrt{s}$ = 1800 GeV at the TEVATRON have been performed with the
collider detector CDF \cite{X1}. Here,
the dependences of the jet shape $\rho$ on the transverse energy $E_T$
and on $r$ were studied
for jets produced in the central region \cite{X1} and compared
with theoretical calculations. The measured jets at CDF
have sufficiently high $E_T$
of order 100 GeV, so that one can assume that gluon emission
effects are much more important than long-distance fragmentation
processes or soft interactions with ``spectator'' partons in
determining the jet shape function $\rho(r,R,E_T,\eta)$.
This means that at sufficiently high energies, the shape of the jet
should be calculable by perturbative QCD alone ignoring
fragmentation effects and other soft interactions.
Such calculations of $\rho$ based on $\alpha_s^3$
finite-order perturbative QCD for $p\overline{p}$
collisions were first performed by Ellis, Kunszt and Soper (EKS)
\cite{X2} and the results were compared to the CDF data \cite{X1}. \\

The production of high transverse energy jets by quasi-real
photons on protons (low $Q^2$
$ep$ scattering) has some similarity to jet production in $p\overline{p}$
collisions. As is well known, two mechanisms contribute to the
photoproduction of jets at large transverse energy. The photon
can either interact directly or via its quark and gluon content,
the resolved contribution, with the partons originating from the
proton. The resolved photoproduction dominates at HERA for lower $E_T$
and positive rapidities $\eta$ (we assign $\eta > 0$
in the direction of the incoming proton). Near the maximum of the
rapidity distribution $(\eta\simeq 2)$,
the resolved and direct parts become comparable at $E_T\simeq 30$
GeV. Due to the dominance of the resolved component, jet shapes in low
$Q^2$ $ep$ collisions should be rather similar to jet shapes in
$p\overline{p}$ jet production.
Next-to-leading order (NLO) QCD calculations of jet shapes at HERA
for the resolved component have been presented some time ago
\cite{X3}. Their similarity to jet shapes in $p\overline{p}$
collisions has been studied in \cite{X4} showing that due to a simple
scaling behavior jet shapes for different c.m.~energies and
reactions can be correlated. This scaling behavior was used by the H1
collaboration to compare jet profiles as measured in $\gamma p$ reactions
\cite{X21} with the jet profiles obtained from $p\overline{p}$ data \cite{X20}.
\\

Recently, the ZEUS collaboration presented their first results for jet
shape functions $\rho$ in $ep$ collisions \cite{X5}. They studied jet shapes
$\rho(r,R,E_T,\eta)$ in inclusive single-jet
and dijet photoproduction for jets with
$E_T > 14$ GeV, $-1 < \eta < 2$, and $R = 1$ as a function of $E_T$, $\eta$,
and the inner cone radius $r$. These measurements show that $\rho$
changes appreciably with $\eta$ and $E_T$
for fixed $r$. In this work we shall calculate $\rho$
for the experimental conditions of the ZEUS measurements in NLO
including direct and resolved contributions to find out whether
the behavior of $\rho$ with varying $\eta$ and $E_T$
found by ZEUS can be explained by fixed-order perturbative
calculations. In particular we shall investigate how the merging
of partons in relation to the merging procedure in
the experimental analysis can be exploited to obtain a
satisfactory description of the experimental results.
With the same attitude we shall investigate jet shapes in
$p\overline{p}$ collisions at the TEVATRON and compare our results with
measurements of CDF \cite{X1} and D0 \cite{X6} in different
rapidity and transverse energy regions. \\

The outline of the paper is as follows. In section 2 we discuss in detail the
merging procedure in the theoretical calculations and in the experimental
analysis of low $Q^2$ $ep$ collisions. In section 3 we compare our numerical
results with the recent ZEUS data. The $p\overline{p}$
jet shapes are considered in section 4. The results of the
calculation and the comparison with CDF and D0 data are given.
Section 5 summarizes our results and ends with some concluding remarks.

\section{Jet Shapes in Low $Q^2$ $ep$ Collisions}
\subsection{Theoretical Input}
We consider positrons with energy
$E_e = 27.5$ GeV colliding with protons of energy
$E_p = 820$ GeV. The photon emission is described by the
Weizs\"acker-Williams equivalent photon approximation with a
formula as in \cite{X7} with
$Q_{\max}^2 = 4$ GeV$^2$
and $0.2 < y < 0.85$, where
$y = E_{\gamma}/E_e$
is the fraction of the initial positron energy transferred to the
photon, in accordance with the conditions of the ZEUS
measurements \cite{X5}. For the resolved contribution we need
the parton densities of the photon. We take them from the work of
Gl\"uck, Reya, and Vogt \cite{X8} converted to the $\overline{\mbox{MS}}$
scheme. The parton distributions of the proton are taken from CTEQ4M
in the $\overline{\mbox{MS}}$ scheme \cite{X9}. Both are NLO parametrizations.
The details of these structure functions are not essential, since $\rho$
is a ratio of cross sections. For
$\alpha_s(\mu)$
we employ the one-loop formula with
$N_f = 4$
and with
$\Lambda$ taken from the proton density
$(\Lambda_{\overline{\rm MS}}^{(4)} = 296$ MeV).
We decided on the one-loop expression for
$\alpha_s$
since the jet shape is calculated only to lowest non-trivial order
${\cal O} (\alpha_s)$.
The renormalization scale
$\mu$
is set equal to the factorization scales and is put equal to
$E_T$. \\

For calculating the jet profile
$\rho$
we need a jet definition. We adopt the cone algorithm of the
Snowmass convention \cite{X10}. According to this definition a
jet is defined as transverse energy
$E_T$
deposited in a cone of radius $R$ in the rapidity-azimuthal angle
plane. The jet axis is determined from
\begin{equation}
 \eta_J =\sum_{i\in{\rm cone}} E_{T,i}\eta_i/E_T,
\end{equation}
\begin{equation}
 \phi_J =\sum_{i\in{\rm cone}} E_{T,i}\phi_i/E_T
\end{equation}
with the transverse energy calculated from
\begin{equation}
 E_T = \sum_{i\in{\rm cone}} E_{T,i}.
\end{equation}
A parton with kinematic variables
$(\eta_i,\phi_i)$
is included in the jet if the condition
\begin{equation}
\sqrt{(\eta_i-\eta_J)^2+(\phi_i-\phi_J)^2} \leq R
\end{equation}
is satisfied. Since we have only up to three partons in the final state,
not more than two partons can be combined into one jet. Since the
jet shape
$\rho$
is fully determined from the
$2 \rightarrow 3$
parton transition cross section we have the situation that the
final state consists either of three or of two jets, where one of
the jets resulted from the recombination out of two partons in
the 3-parton final state. \\

As is well known this jet definition
has two problems: double counting and merging. The condition (4)
is equivalent to
\begin{equation}
 \sqrt{(\eta_i-\eta_j)^2+(\phi_i-\phi_j)^2} \leq \frac{E_{T_i}+E_{T_j}}
 {\max(E_{T_i},E_{T_j})} R,
\end{equation}
i.e.~if the parton angles of partons $i$ and $j$ satisfy this condition,
they are counted as one combined jet with transverse energy
$E_{T_i}+E_{T_j}$,
but not as two smaller jets with energies
$E_{T_i}$
and
$E_{T_j}$.
For the case that
\begin{equation}
 R < \sqrt{(\eta_i-\eta_j)^2+(\phi_i-\phi_j)^2} \leq \frac{E_{T_i}+E_{T_j}}
 {\max(E_{T_i},E_{T_j})} R,
\end{equation}
which occurs, for example, for
$E_{T_i}=E_{T_j}=E_T/2$,
the two partons $i$ and $j$ might with some justification count also
as separate jets. Therefore one has the choice to count only the
combined jet $\{i$ and $j\}$ or to count $\{i\}$, $\{j\}$, and $\{i$ and $j\}$
as
separate jets which would contribute to the inclusive single-jet
cross section. We follow EKS \cite{X11} and count only the combined
jet in this case. This way we avoid any double counting of jets. In
practice it is found, however, that the jet shape
$\rho$
is unaffected by the double counting issue of jet recombination except
at
$r\simeq 0$ \cite{X3}. \\

The issue of merging is much more severe. It is related to the
problem whether the cone algorithm used in the theoretical NLO
calculation matches the jet definition in the experimental analysis.
This problem cannot be solved easily since in the experimental
measurement of
$\rho$
one starts with the observation of hadrons which are recombined
into protojets with a cone algorithm similar to the Snowmass definition.
These protojets are recombined further until the conditions of the
algorithm which include merging of overlapping cones are satisfied.
The details of the algorithm used by the ZEUS collaboration will
be described later. It is clear, however, that without further
investigations it is not possible to decide that cone algorithms
used for three parton final states match the cone algorithms used
in the recombination of final state hadrons in the ZEUS analysis. \\

The theoretical jet algorithm we are using will merge two partons
into a single jet whenever the condition (5) is satisfied. This
includes also the configuration when two partons with equal
transverse energy
$E_{T_i}=E_{T_j}=E_T/2$
are just $2R$ apart. In the calculation this is counted as one jet
of transverse energy
$E_T$
with a cone centered between the two partons. This means that for
this case the
$E_T$
of the jet is distributed at the edge at the cone and not in the
vicinity of the center. Whether such a configuration will be
counted as a single jet also in the experimental analysis will,
of course, depend on the details of the jet algorithm used in the
experiment. This problem was recognized by EKS \cite{X2} when they
calculated jet shapes
$\rho$
for
$p\overline{p}$
collisions and compared them with the CDF data \cite{X1}. To simulate
the merging and other conditions of the CDF cone algorithm in a
simple way they added the extra constraint, that two partons, $i$ and $j$,
when separated by more than $R_{\rm sep} \leq 2R$, i.e.~when
\begin{equation}
 \sqrt{(\eta_i-\eta_j)^2+(\phi_i-\phi_j)^2} \geq R_{\rm sep},
\end{equation}
are no longer merged into a single jet. With this additional
constraint the fraction of
$E_T$
near the edge of the cone is reduced. EKS also found that
$R_{\rm sep} = 1.3R$ describes the jet shape
$\rho$
as a function of $r$ for
$E_T = 100$ GeV jets as measured by CDF very well. This means that for
simulating the CDF cone algorithm the $R_{\rm sep}$ constraint with
$R_{\rm sep} = 1.3R$ should be introduced on the parton level. Of course,
this value for $R_{\rm sep}$ accounts only for the ideal situation of high
$E_T$
jets at CDF with very small fragmentation, underlying event, and
other effects, which might distort the relation of the CDF cone
algorithm to the algorithm employed on the parton level. From this
we conclude that the jet merging in the CDF analysis can be
simulated on the parton level with the parameter
$R_{\rm sep} \leq 2R$
which reduces the merging of partons at the cone edge. For
$E_T \simeq 100$ GeV jets $R_{\rm sep} \simeq 1.3R$. For lower $E_T$'s
we expect additional broadening of the jets which we might simulate
with $R_{\rm sep}$ values slightly larger than $1.3R$. Therefore we consider
$R_{\rm sep}$ a parameter which is not fixed completely and depends on the
kinematical parameters of the final state jets. Before we decide
which $R_{\rm sep}$ values should be assumed for the photoproduction of jets
at HERA we shall describe the jet algorithm employed in the ZEUS
analysis.

\subsection{Jet Algorithm in the ZEUS Analysis}
In the ZEUS jet search and the measurement of the jet shape $\rho$
\cite{X5} a cone algorithm \cite{X10} is used to construct jets
on the basis of the energy depositions in the uranium-scintillator
calorimeter (CAL) cells in both data and simulated events
and also from the final state hadrons in the simulated events.
Whether hadrons or CAL cells are considered is not essential. The
following steps are taken to construct jets. CAL cells are
considered with their $\eta$ and $\phi$
determined from the unit vectors joining the vertex of the
interaction and the geometric centers of the cells. In the first
step, each CAL cell with a transverse energy above 0.3 GeV is
considered as a seed for the search. These seeds are combined if
their distance in $\eta-\phi$ space
$R = \sqrt{(\Delta\eta)^2+(\Delta\phi)^2}\leq 1$.
The cone radius $R = 1$ is drawn around each seed and the CAL cells
within that cone are combined to form a cluster. The axis of the
cluster is defined according to the Snowmass conventions \cite{X10}.
This yields $\eta_{\rm cluster}$ and $\phi_{\rm cluster}$
which follow from the transverse energy weighted mean
pseudorapidity and azimuthal angle of all the CAL cells belonging
to this cluster (see (1), (2), and (3)). A new cone of radius $R = 1$
is then drawn around the axis of the resulting cluster. All cells
with their geometric center inside the cone are used to recalculate
a new cluster axis. The procedure is iterated until the content of the cluster
is stable. Up to this point the algorithm is very similar
to the jet definition at the parton level which we use to produce
our predictions for comparison with the experimental data. Of course,
with only three partons in the final state, the procedure is much
simpler. Iterations are not needed and the possibilities for
merging are very limited, since maximally only two partons can be
merged. \\

The difference to the theoretical algorithm occurs in the treatment
of overlapping clusters. In the ZEUS analysis two clusters are merged
when the common transverse energy exceeds 75\% of the total
transverse energy of the cluster with the
smallest transverse energy. Otherwise two different clusters are
formed and the overlapping cells are assigned to the nearest
cluster. Finally a cluster is called a jet if the corrected $E_T$
exceeds 14 GeV. Concerning the rapidity only jets with $\eta$
in the range $-1<\eta<2$ are selected. \\

We emphasize that in this experimental procedure of defining jets there are
several steps that have no analogy in the theoretical jet definition with only
three partons in the final state. First, in the experimental definition
iterations are needed to define stable clusters. Second, there is the
merging of two clusters depending on the shared energy. In the theoretical
definition there is neither an iteration nor shared energy. The latter
would only occur in a theoretical analysis with at least four partons in the
final state, i.e. if we went to one order higher in the QCD calculation.
In this case we have NNLO two-jet, NLO three-jet, or LO four-jet production.
In ${\cal O} (\alpha_s^3)$, where there is only NLO two-jet and LO three-jet
production, merging of partons occurs only once, so that iterations and
overlapping cones do not arise. \\

The energy sharing and merging of overlapping clusters in the
ZEUS analysis is the same procedure as used by the CDF collaboration
to define jets in $p\overline{p}$ collisions. In both measurements, jets are
constructed with the choice $R=1$. Therefore we expect that the jet
definition in the ZEUS analysis corresponds to $R_{\rm sep}\simeq 1.3R$
in the merging of partons for the theoretical jet
algorithm at least for the ideal case of large $E_T$
and/or jet production in the backward direction. For negative $\eta$'s
the jet production is a superposition of nearly equal
contributions from direct and resolved production with
$x_{\gamma}\simeq 1$ ($x_{\gamma}$
is the fraction of the initial positron energy transferred to the
photon). In this region there is little disturbance of the jets
through additional interaction with the remnants. In the region
$x_{\gamma}\geq 0.75$
the measured transverse energy flow around the jet axis is
described reasonably well without extra multiple interactions
included in the Monte Carlo \cite{X12}. This is already the case
for rather low $E_T$ but improves the larger the $E_T$ is.
Of course, the introduction of the $R_{\rm sep}$ parameter into the theoretical
jet definition can only be a phenomenological device to model the much
more involved experimental jet definition. \\

Further information on $R_{\rm sep}$ can be gained from a study of
experimental and theoretical jet definitions for photoproduction
\cite{X13}. In this study inclusive dijet cross sections
d$\sigma/$d$\overline{\eta}$ with $E_T > 6$ GeV and
$|\eta_1-\eta_2| \leq 0.5$
for direct and resolved photoproduction as a function of the average
pseudorapidity of two observed jets $\overline{\eta}=(\eta_1+\eta_2)/2$
were compared for three jet definitions denoted EUCELL, PUCELL and
KTCLUS. The first two definitions are cone algorithms whereas KTCLUS
is a cluster algorithm. PUCELL is identical to the jet definition
described above. KTCLUS is equivalent to the cone algorithm with
$R_{\rm sep} = 1$. The inclusive dijet cross sections vary with the jet
definitions. EUCELL corresponds to $R_{\rm sep}$ = 2 and yields the largest
cross sections. Calculations with the HERWIG Monte Carlo \cite{X14}
reveal that the cross sections with PUCELL are larger (almost equal)
compared to the KTCLUS cross section for resolved (direct)
production. The resolved (direct) cross sections are again defined
with the
$x_{\gamma} < 0.75$ $(x_{\gamma} > 0.75)$
cut. By comparing with the EUCELL $(R_{\rm sep} = 2)$ cross section
we conclude that the resolved PUCELL cross section corresponds
to a cross section which we would obtain in the NLO calculation
with $R_{\rm sep} \simeq 1.4R$.
This is consistent with recent measurements of these dijet cross
section with different jet definitions \cite{X15}. Here for the
enriched resolved
$\gamma$
sample
$(0.3 < x_{\gamma} < 0.75)$
the cross section for PUCELL is approximately 30\% larger than the
KTCLUS cross section. Both cross sections have appreciable errors,
so that definite conclusions about equivalent $R_{\rm sep}$ values are
difficult to obtain. Furthermore for the measurement of jet shapes
only inclusive single jets are considered so that it is not justified to read
off the exact values for $R_{\rm sep}$ from the dijet analysis described above.
However, we are confident that for the PUCELL algorithm
$R_{\rm sep} \geq 1.3R$
must be assumed when applied to the calculation of jet shapes where
the resolved component is dominant. From earlier calculations
\cite{X2, X4} we know that
$\rho$
depends sensitively on $R_{\rm sep}$. The larger $R_{\rm sep}$ is the broader
the jets are. \\

The jet shapes for the reaction
$e+p \rightarrow \mbox{jet}+X$
have been measured for jets with $E_T>14$ GeV
integrated over four non-overlapping $\eta$
regions $(-1<\eta<0,~0<\eta<1,~1<\eta<1.5$, and $1.5<\eta<2)$
and are presented also in four $E_T$ regions
$(14~\mbox{GeV}<E_T<17~\mbox{GeV},~17~\mbox{GeV}<E_T<21~\mbox{GeV},
21~\mbox{GeV}<E_T<25~\mbox{GeV},$ and $E_T>25~\mbox{GeV})$
integrated always over the same $\eta$ interval $-1<\eta<2$.
It would not be reasonable to insist that the jet shapes $\rho$
for these different $\eta$ and $E_T$
regions should be calculated always with the same $R_{\rm sep}$ parameter.
We expect $R_{\rm sep} \simeq 1.3R$ for the $\eta$ interval $-1 < \eta < 0$
where direct and resolved production contribute in nearly equal
amounts and it is known that single-jet production in this region
is dominated by
$x_{\gamma} \simeq 1$.
Away from the backward (direct) region we expect increasingly broader jets
the more we approach the forward direction
$\eta \rightarrow 2$.
We know from earlier investigations that the single inclusive jet
cross section as a function of
$\eta$
for
$E_T \geq E_T^{\min}$
as measured by ZEUS \cite{X16} does not agree with the NLO
prediction in the region $\eta>1$
\cite{X17}, the experimental cross sections are larger than the
theoretical ones. This excess of the cross section from $\eta>1$
could be simulated in the PYTHIA Monte Carlo \cite{X18} by adding
multiple interactions. These multiple interactions, which apply only
to resolved processes, consist of interactions between partons
in the proton and photon remnants calculated as LO processes
and generated in addition to the primary hard scattering. These
multiple interactions lead to an energy flow outside the core of the jet
due to a possible underlying event. In \cite{X5} it was found
that these multiple interaction effects on the jet shapes are small
in the region
$-1<\eta<1$,
but increase gradually with
$\eta$
in the region
$\eta > 1$,
where an improved description of the data is obtained. In our NLO
calculation we shall describe this excess of energy outside of the
core of the jet with a gradual increase of $R_{\rm sep}$ towards
$\eta=2$.

\section{Results for Jet Shapes in $ep$ Collisions}
As already introduced in section 1, for a sample of jets of
transverse energy
$E_T$,
defined with a cone radius $R$, the jet shape
$\rho(r,R,E_T,\eta)$
is the average fraction of the jet's transverse energy that lies
inside an inner cone with radius $r < R$, which is concentric with the
jet defining cone. Then the quantity
$1-\rho$
stands for the fraction of
$E_T$
that lies in the cone segment between $r$ and $R$. This is given, up to
higher order terms in $\alpha_s$, by
\begin{equation}
 1-\rho(r,R,E_{T,J},\eta) = \frac{\int\mbox{d}E_T E_T\mbox{d}\sigma(e+p
 \rightarrow e'+3~\mbox{partons}+X)/\mbox{d}E_T}{E_{T,J}
 \sigma(E_{T,J})_{\rm LO}},
\end{equation}
where the integral in the numerator is performed over the cone
segment between $r$ and $R$. This quantity is calculated from the
contributing
$2 \rightarrow 3$
parton subprocesses. For $r > 0$ the integration does not include
the collinear singularities which are at $r = 0$. Therefore it can
be computed easily from the
$E_T$
weighted integral of cross sections for
$e+p \rightarrow e' + 3~\mbox{partons}+X$
over the cone segment between $r$ and $R$ normalized to
$E_{T,J}$
times the LO cross section.
$1-\rho$
in (8) is
${\cal O}(\alpha_s)$. \\

\begin{figure}[p]
 \begin{center}
  {\unitlength1cm
  \begin{picture}(13,8.7)
   \epsfig{file=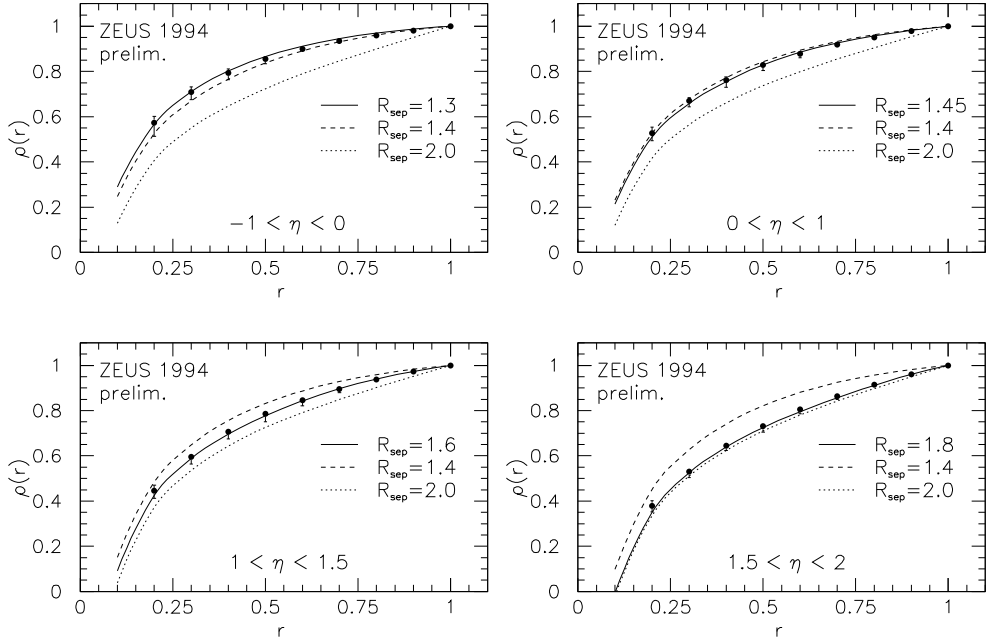,bbllx=520pt,bblly=95pt,bburx=105pt,bbury=710pt,%
           height=13cm,clip=,angle=270}
  \end{picture}}
  \caption{\label{plot1}{\it Jet shape $\rho(r)$ for complete
            single-jet photoproduction integrated over $E_T>14$ GeV and four
            different regions of $\eta$. We compare our results using the
            Snowmass convention with $R=1$ and three different values of
            $R_{\rm sep}$ to preliminary 1994 data from ZEUS.}}
 \end{center}
\end{figure}

From (8) we have calculated $\rho(r,R,E_T,\eta)$ as a function of $r$ for
the four rapidity intervals $-1<\eta<0,~0<\eta<1,~1<\eta<1.5$, and
$1.5<\eta<2$ taking $R = 1$ and integrated over $E_T > 14$ GeV.
Here we identify $E_{T,J}$ in (8) with the $E_T$ of the jet.
For the $R_{\rm sep}$ parameter we have made three choices: (i) $R_{\rm sep}
= 1.4R = 1.4$ (since $R = 1$), which we consider as an average value
appropriate for the PUCELL
algorithm in the ZEUS analysis, (ii) $R_{\rm sep}$ = 2 as the maximal
possible value, and (iii) an optimized $R_{\rm sep}$ value, which is
$R_{\rm sep}$ = 1.3, 1.45, 1.6, and 1.8 for the four rapidity regions.
Our results, plotted in fig.~\ref{plot1},
%
\begin{figure}[p]
 \begin{center}
  {\unitlength1cm
  \begin{picture}(13,8.7)
   \epsfig{file=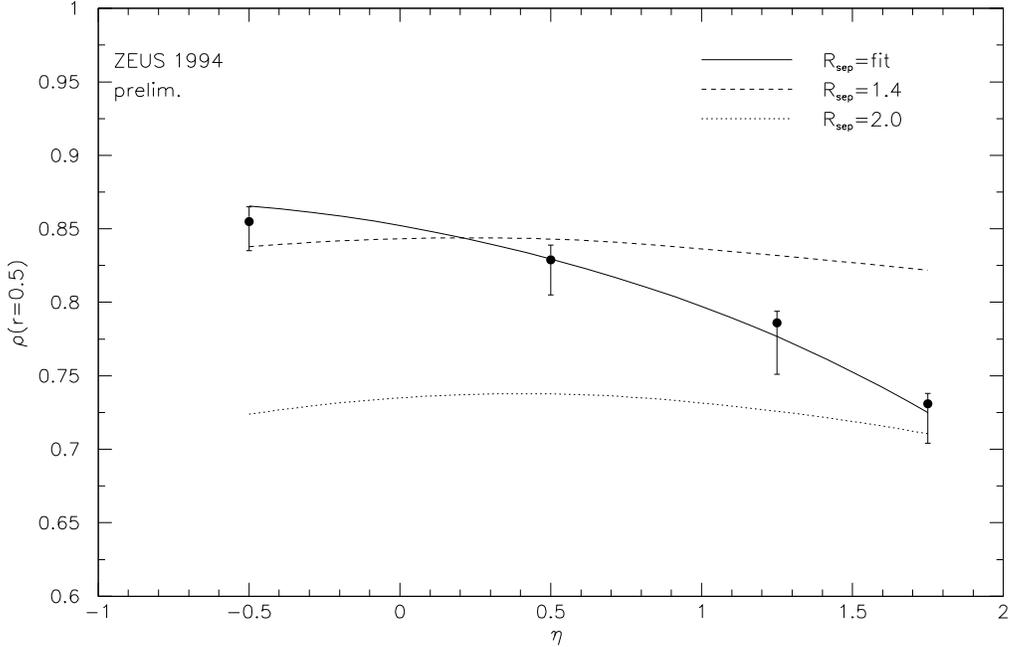,bbllx=520pt,bblly=95pt,bburx=105pt,bbury=710pt,%
           height=13cm,clip=,angle=270}
  \end{picture}}
  \caption{\label{plot2}{\it Jet shape $\rho(\eta)$ for complete
            single-jet photoproduction for $r=0.5$ and integrated over
            $E_T>14$ GeV and the same regions of $\eta$ as
            in the last figure. We compare our results with
            one variable and two fixed
            values of $R_{\rm sep}$ to preliminary 1994 data from ZEUS.}}
 \end{center}
\end{figure}
are compared with the preliminary
data from ZEUS \cite{X5}. We observe, that for fixed $R_{\rm sep}$ the
jet profile
$\rho$
does not depend very much on
$\eta$
except for small $r$. Furthermore
$\rho$
depends sensitively on $R_{\rm sep}$. The jets become broader with
increasing $R_{\rm sep}$. Neither $R_{\rm sep}$ = 2 nor $R_{\rm sep}$ = 1.4
agree with the data in the four
$\eta$
regions. However, this is to be expected according to the discussion
in the previous section. In the first
$\eta$
interval,
$-1<\eta<0$,
the data are very well accounted for with $R_{\rm sep}$ = 1.3, which is the
value we anticipated for the ideal situation with no extra
broadening due to multiple interaction effects with the remnants.
With increasing
$\eta$
the value of $R_{\rm sep}$ must be increased monotonically up to $R_{\rm sep}$
= 1.8
in the fourth
$\eta$
interval,
$1.5<\eta<2$.
Due to the additional broadening from multiple scattering effects
towards
$\eta = 2$
we expect this increase of $R_{\rm sep}$ with increasing
$\eta$.
Since we are unable to predict the $R_{\rm sep}$ values for the four
$\eta$
intervals in advance, the optimized $R_{\rm sep}$'s come just from the fit.
However, we know that $1.3<R_{\rm sep}<2.0$ in the whole
$\eta$ range. This way, with choosing $R_{\rm sep}$ larger than 1.3, we
simulate in addition to the modelling of the experimental jet definition
broadening effects which are not taken into account in the perturbative
calculation and which may arise from multiple interactions or other effects. \\

If we fix $r = 0.5$, the dependence of $\rho(r)$ as a function of $\eta$
over the range $-0.5<\eta<1.75$
is shown in fig.~\ref{plot2}. For fixed $R_{\rm sep}$ the jet shape
$\rho(r=0.5)$ depends only marginally on $\eta$
in contrast to the experimental data \cite{X5}. As to be expected, the
$\eta$ profile is in agreement with the data for the optimized $R_{\rm sep}$
(denoted $R_{\rm sep}$ = fit) given by the full curve in fig.~\ref{plot2}. \\

\begin{figure}[p]
 \begin{center}
  {\unitlength1cm
  \begin{picture}(13,8.7)
   \epsfig{file=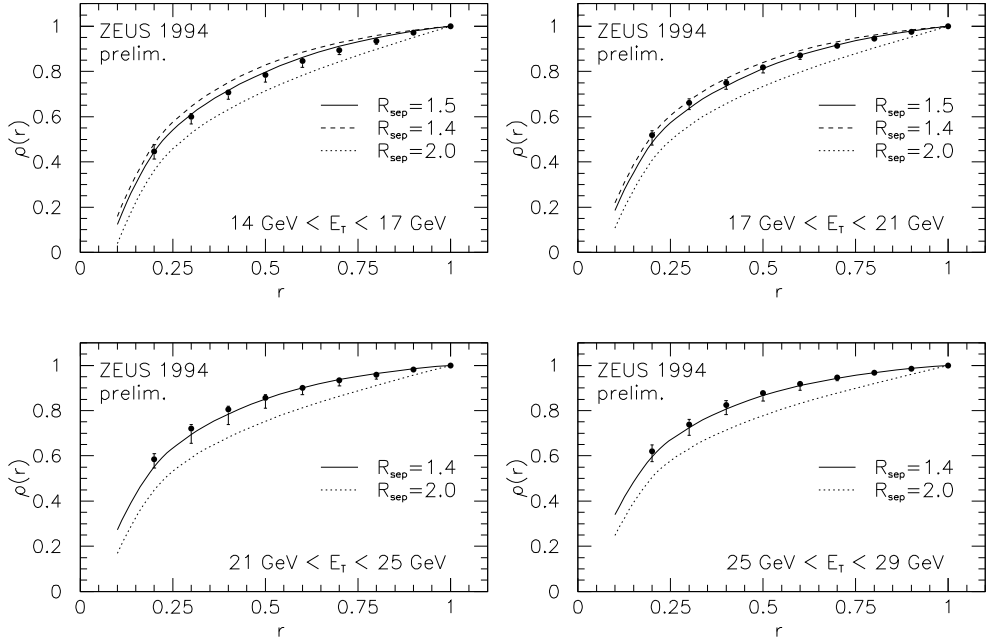,bbllx=520pt,bblly=95pt,bburx=105pt,bbury=710pt,%
           height=13cm,clip=,angle=270}
  \end{picture}}
  \caption{\label{plot3}{\it Jet shape $\rho(r)$ for complete
            single-jet photoproduction integrated over $-1<\eta<2$ and four
            different regions of $E_T$. We compare our results using the
            Snowmass convention with $R=1$ and different values of
            $R_{\rm sep}$ to preliminary 1994 data from ZEUS.}}
 \end{center}
\end{figure}

In the ZEUS jet shape measurement, $\rho$
is also presented as a function of $r$ for four non-overlapping $E_T$
regions: $14~\mbox{GeV}<E_T<17~\mbox{GeV},~17~\mbox{GeV}<E_T<21~\mbox{GeV},~
21~\mbox{GeV}<E_T<25~\mbox{GeV}$, and $25~\mbox{GeV}<E_T<29~\mbox{GeV}$,
integrated over $-1<\eta<2$. Their results are shown in fig.~\ref{plot3}
%
\begin{figure}[p]
 \begin{center}
  {\unitlength1cm
  \begin{picture}(13,8.7)
   \epsfig{file=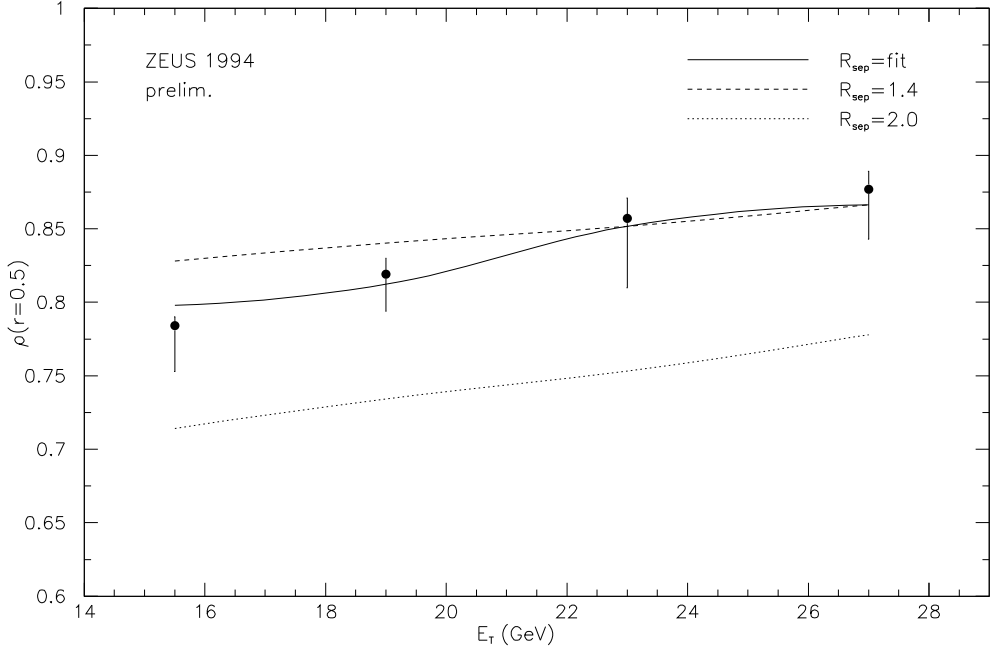,bbllx=520pt,bblly=95pt,bburx=105pt,bbury=710pt,%
           height=13cm,clip=,angle=270}
  \end{picture}}
  \caption{\label{plot4}{\it Jet shape $\rho(E_T)$ for complete
            single-jet photoproduction for $r=0.5$ and integrated over
            $-1<\eta<2$ and the same regions of $E_T$ as
            in the last figure. We compare our results with
            one variable and two fixed
            values of $R_{\rm sep}$ to preliminary 1994 data from ZEUS.}}
 \end{center}
\end{figure}
compared to theoretical
calculations with $R_{\rm sep}$ = 2, $R_{\rm sep}$ = 1.4 and an optimized
$R_{\rm sep}$
which varies between 1.4 and 1.5. We see that $R_{\rm sep}$ = 1.4 yields in
average a fairly good description of the data points over the whole $E_T$
region. Only in the first two $E_T$
regions we need a larger value $R_{\rm sep}$ = 1.5. From this comparison
we conclude that $R_{\rm sep}$ increases when $E_T$
decreases. \\

This is seen also when we plot $\rho(r=0.5)$ as a function of $E_T$
for $15.5~\mbox{GeV}<E_T<27~\mbox{GeV}$ in fig.~\ref{plot4}
and compare with the data from \cite{X5}. Whereas for
fixed $R_{\rm sep}$ = 1.4 the jet shape
$\rho(r=0.5)$
decreases only little with decreasing
$E_T$
the experimental point at
$E_T = 15.5$ GeV lies below the curve showing that here $R_{\rm sep}$ is larger
than 1.4, i.e.~the jet broadening with decreasing
$E_T$
\cite{X3} is somewhat stronger than predicted for a fixed $R_{\rm sep}$. \\

\begin{figure}[p]
 \begin{center}
  {\unitlength1cm
  \begin{picture}(13,8.7)
   \epsfig{file=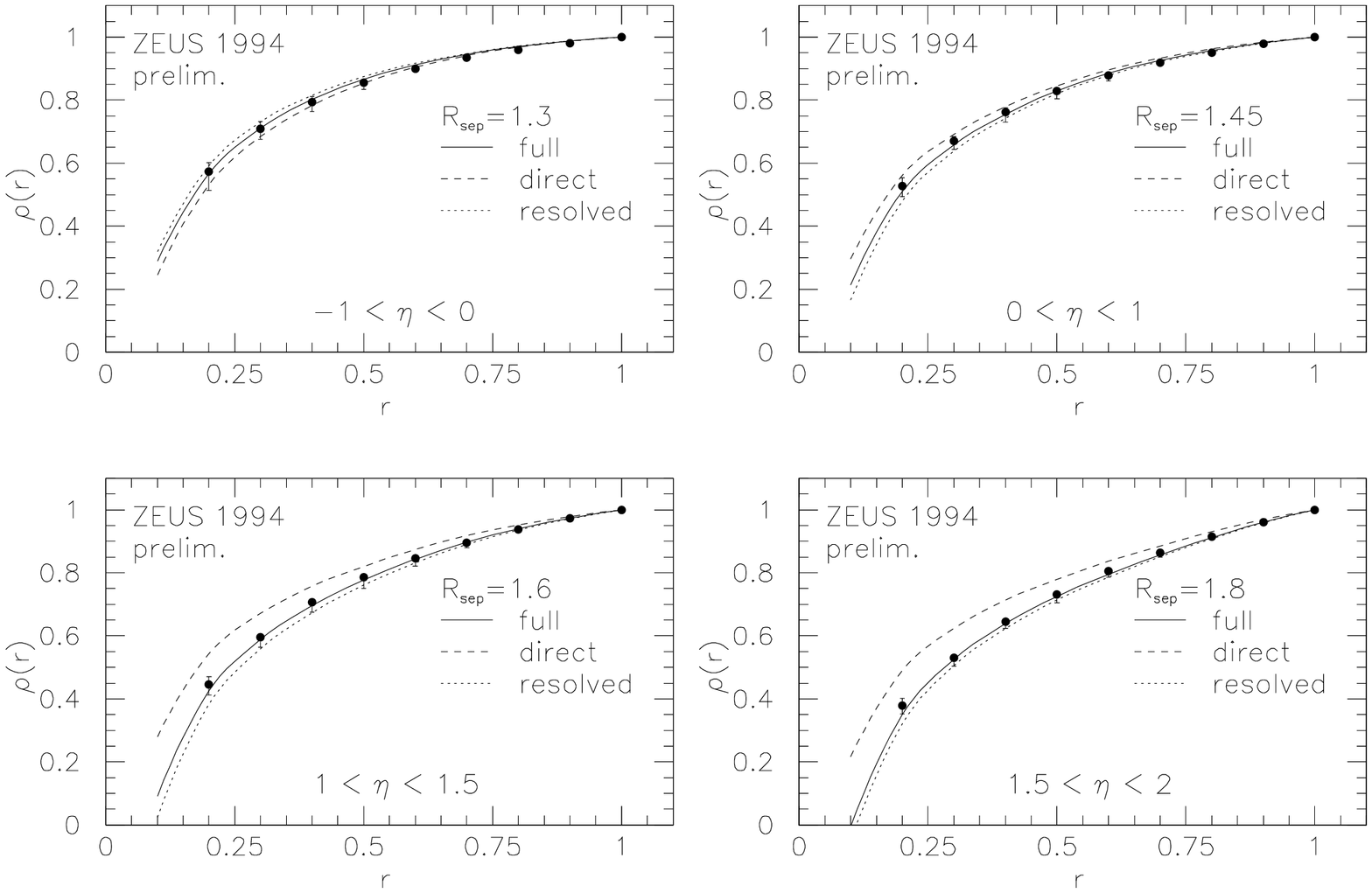,bbllx=520pt,bblly=95pt,bburx=105pt,bbury=710pt,%
           height=13cm,clip=,angle=270}
  \end{picture}}
  \caption{\label{plot5}{\it Jet shape $\rho(r)$ for full, direct, and resolved
            single-jet photoproduction integrated over $E_T>14$ GeV and four
            different regions of $\eta$. We compare our results using the
            optimal value of $R_{\rm sep}$ to preliminary 1994 data from
            ZEUS.}}
 \end{center}
\end{figure}

In fig.~\ref{plot5},
%
\begin{figure}[p]
 \begin{center}
  {\unitlength1cm
  \begin{picture}(13,8.7)
   \epsfig{file=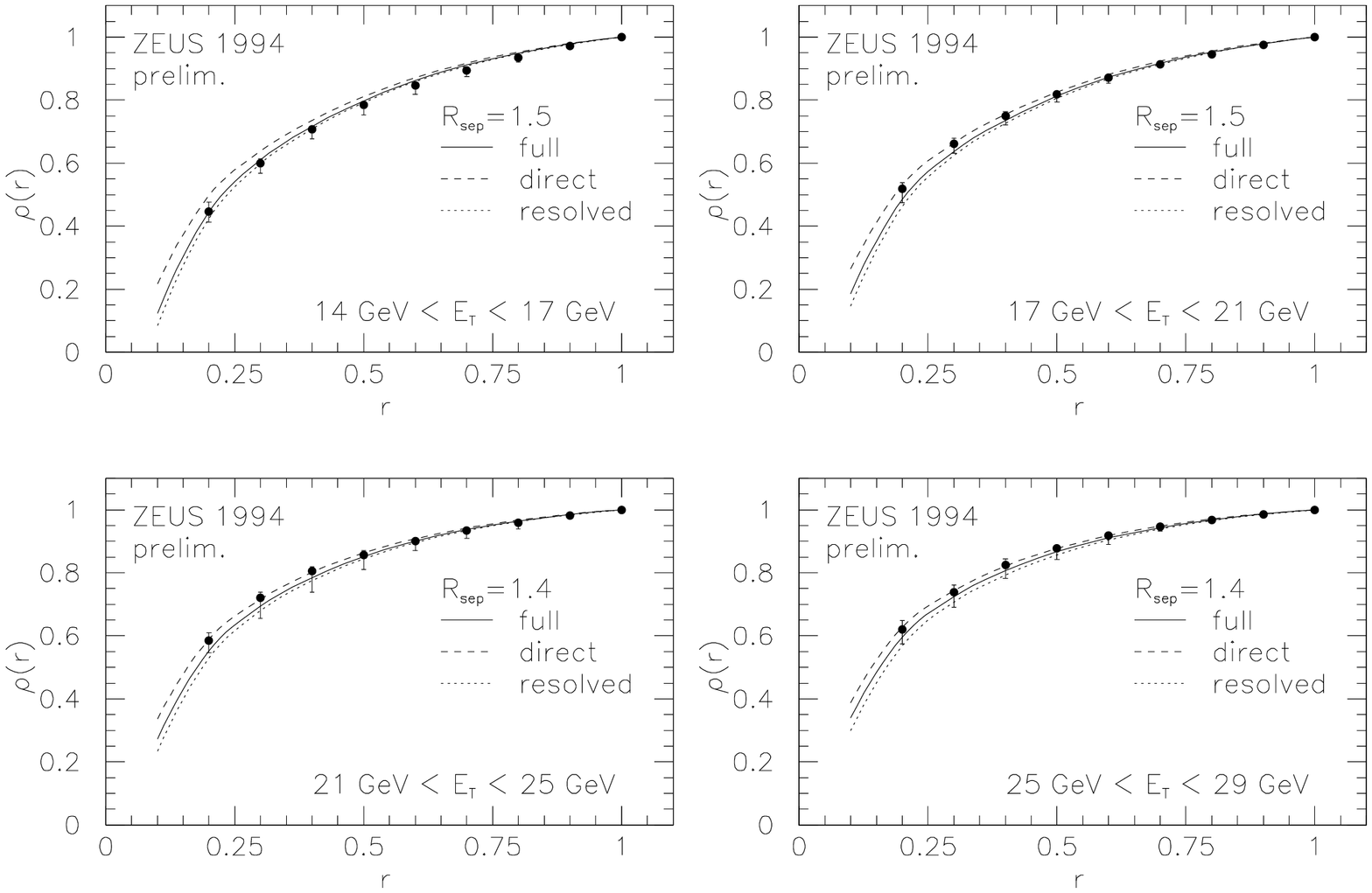,bbllx=520pt,bblly=95pt,bburx=105pt,bbury=710pt,%
           height=13cm,clip=,angle=270}
  \end{picture}}
  \caption{\label{plot6}{\it Jet shape $\rho(r)$ for full, direct, and resolved
            single-jet photoproduction integrated over $-1<\eta<2$ and four
            different regions of $E_T$. We compare our results using the
            optimal value of $R_{\rm sep}$ to preliminary 1994 data from
            ZEUS.}}
 \end{center}
\end{figure}
we show the jet shapes for the four $\eta$ intervals and $E_T>14~\mbox{GeV}$
as in fig.~\ref{plot1} with optimized $R_{\rm sep}$ for the direct (dashed
lines) and resolved processes (dotted lines) separately together with the
curves, where the direct and resolved contributions are
superimposed (full lines). It is known that at low $E_T$
the resolved component dominates. Therefore the full curves lie very
near to the resolved curves. The jet shapes for the direct process
are, except for the first
$\eta$
region
$-1<\eta<0$,
narrower than those of the resolved process leading to slightly
narrower jets when both contributions are superimposed. In the
region
$\eta<0$
the role of direct and resolved contributions are reversed. Here
the direct jet shape is slightly broader than for the resolved
process. \\

The same plots for the jet shapes as a function of $r$
for the four
$E_T$
intervals as in fig.~\ref{plot3} are exhibited in fig.~\ref{plot6}.
Here all events for
$-1<\eta<2$
are included. For
$\rho$
at the smaller $r$ one can see that the direct contribution becomes
more important with increasing
$E_T$
as we expect. \\

\begin{figure}[ht]
 \begin{center}
  {\unitlength1cm
  \begin{picture}(13,8.7)
   \epsfig{file=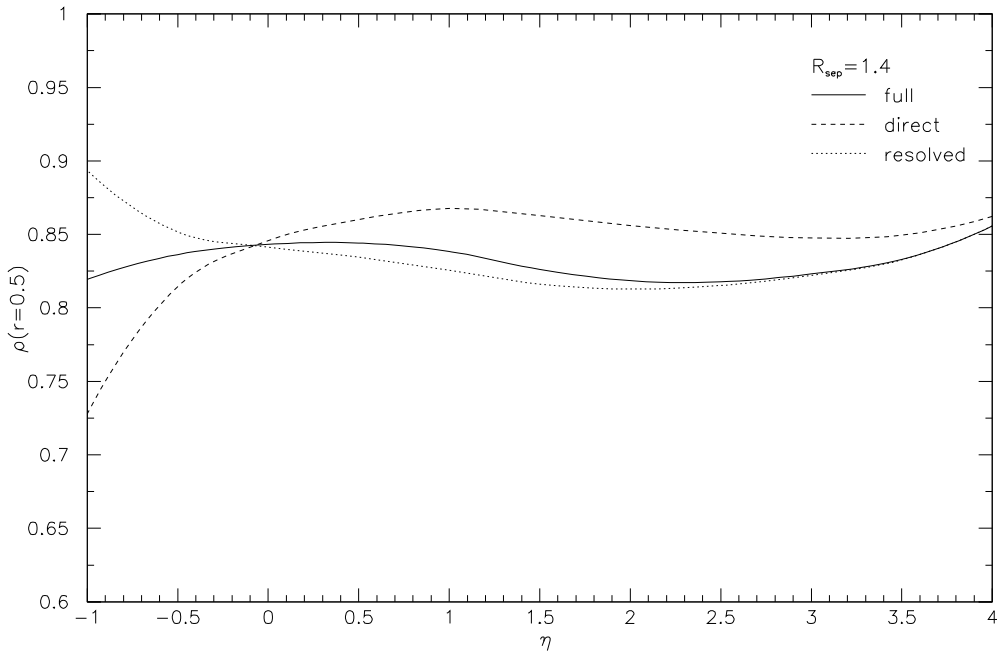,bbllx=520pt,bblly=95pt,bburx=105pt,bbury=710pt,%
           height=13cm,clip=,angle=270}
  \end{picture}}
  \caption{\label{plot7}{\it Jet shape $\rho(\eta)$ for full, direct, and
            resolved single-jet photoproduction for $r=0.5$, $E_T > 14$ GeV,
            and $R_{\rm sep}
            =1.4$. Direct and resolved contributions have different shapes and
            a crossing point near $\eta\simeq 0$.}}
 \end{center}
\end{figure}

From the comparison in figures~\ref{plot5} and \ref{plot6}
it would be difficult to draw
any conclusions that the direct contribution is needed to explain
the data since the general shape of $\rho$
as a function of $r$ is so similar. For this we need more detailed
information, for example the $\eta$ dependence of $\rho$
for fixed $r$ and fixed $E_T$.
This has been calculated in \cite{X3} for the resolved contribution.
The result was that $\rho(r=0.5)$ for fixed $E_T$
and $R_{\rm sep}$ = 2 as a function of $\eta$
had a minimum, i.e. a concave shape. In fig.~\ref{plot2},
$\rho(r = 0.5)$, which has contributions from direct and resolved
processes, has a convex shape with a slight maximum in the considered $\eta$
range. This occurs for both $R_{\rm sep}$ fixed to $R_{\rm sep}$ = 1.4 and
$R_{\rm sep}$ = 2. This different shape of $\rho$ for fixed $r$ and $E_T$
must be characteristic for the direct component. This is indeed
the case. In fig.~\ref{plot7},
we have plotted $\rho(r = 0.5)$ for
$E_T > 14$ GeV and $R_{\rm sep}$ = 1.4 as a function of $\eta$
for the direct and the resolved contribution separately and also
for the superposition of both. As to be expected the resolved
contribution of $\rho$
has the concave shape, and the direct contribution is convex at least for
$\eta<2$. The total contribution is convex in the region $-1<\eta<2$
in agreement with the result in fig.~\ref{plot2}. It would be interesting
to see the increase of $\rho(r = 0.5)$ after the minimum at $\eta \simeq 2.0$
towards increasing $\eta$.
In practice this will be difficult, since for increasing $\eta$,
$R_{\rm sep}$ must be increased (except at large $E_T$),
which leads to a decrease of $\rho (r = 0.5)$ with increasing $\eta$
as shown in fig.~\ref{plot2}. It is interesting, however, that in the backward
direction the shape of $\rho$ as a function of $\eta$ is determined by the
$\rho$ dependence of the direct component, although the resolved
component dominates for $\eta > 0$. The crossing point is near $\eta \simeq 0$.
For $\eta < 0$ both components seem to contribute equally. \\

From the comparison of the jet shapes as measured by ZEUS with our
NLO calculations we conclude that the jet shapes can very well be
explained with a varying $R_{\rm sep}$ parameter describing either
additional jet broading effects or/and any possible changes of the
merging conditions with respect to the jet merging in the
theoretical calculations. In particular we have found that for
$\eta<0$,
$R_{\rm sep}$ = 1.3 as we expect it for the PUCELL algorithm and $R_{\rm sep}$
increases up to $R_{\rm sep}$ = 1.8 in the forward direction (see
fig.~\ref{plot2}).
As a function of
$E_T$
the parameter $R_{\rm sep}$ changes only slightly, when all events for the
whole
$\eta$
range are included
$-1 < \eta < 2$.
$R_{\rm sep}$ increases in this case with decreasing
$E_T$
(see fig.~\ref{plot4}) in accordance with the fact that additional jet
broading effects become more important the lower
$E_T$
is. In the next section we shall see whether these findings can
help to describe the jet shapes in
$p\overline{p}$
collisions as measured by the CDF and D0 collaborations.

\section{Results for Jet Shapes in $p\overline{p}$ Collisions}
\subsection{Comparison with CDF Data}
Quite some time ago the CDF collaboration \cite{X1} has presented
data for the jet shape
$\rho(r,R,E_T,\eta)$
as a function of $r$ for three
$E_T$
intervals
$40~\mbox{GeV}<E_T<60~\mbox{GeV},~65~\mbox{GeV}<E_T<90~\mbox{GeV}$,
and
$95~\mbox{GeV}<E_T<120~\mbox{GeV}$
and rapidities in the central region
$0.1\leq |\eta|\leq 0.7$.
The laboratory system at the TEVATRON is the c.m.
$p\overline{p}$
system. These data were compared with the NLO results of EKS
\cite{X2}. For
$E_T = 100$ GeV, i.e. for the data in the third
$E_T$
bin they found better agreement with the data with $R_{\rm sep}$ = 1.3
instead of the original choice $R_{\rm sep}$ = 2. The measurements by CDF
and the NLO calculations were done with $R = 1$. \\

The energy sharing and merging of overlapping clusters and all
other jet defining conditions in the CDF analysis are the same as
described in subsection 2.2 for the ZEUS analysis. This means
$R_{\rm sep}$ = 1.3 is the right choice in the theoretical calculation
for the treatment of overlapping cones. In our calculations we left
$R_{\rm sep}$ to be free to obtain the best description of the CDF data.
The parton distributions of the proton, which we need as input, are
the same as in the previous section. Also
$\alpha_s$
is calculated in the same way as for the $ep$ jets. \\

\begin{figure}[p]
 \begin{center}
  {\unitlength1cm
  \begin{picture}(13,8.7)
   \epsfig{file=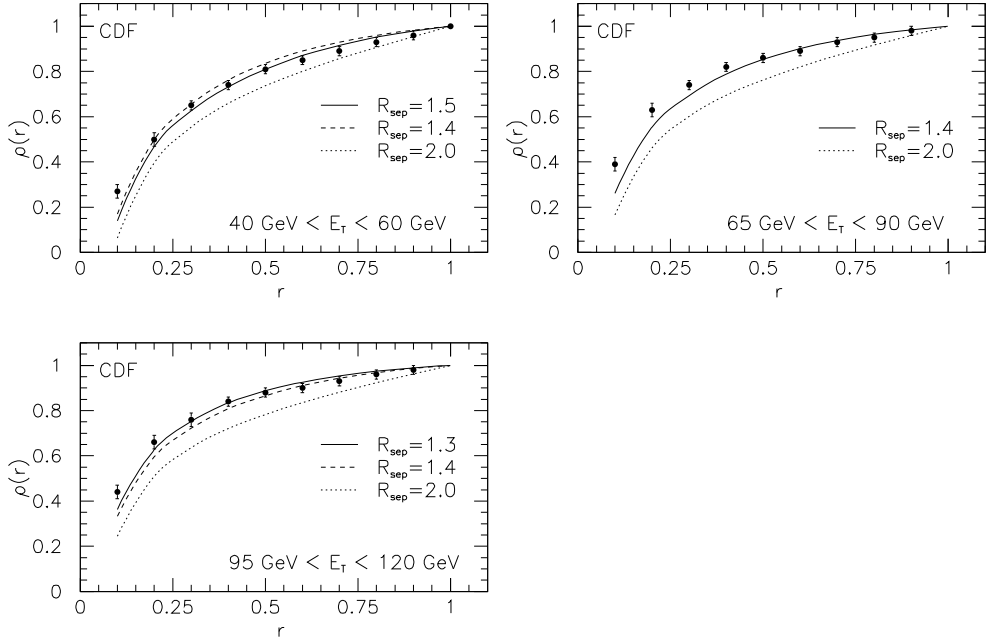,bbllx=520pt,bblly=95pt,bburx=105pt,bbury=710pt,%
           height=13cm,clip=,angle=270}
  \end{picture}}
  \caption{\label{plot8}{\it Jet shape $\rho(r)$ for single-jet production in
            $p\overline{p}$ collisions integrated over $0.1\leq |\eta|\leq 0.7$
            and three different regions of $E_T$. We compare our results using
            the Snowmass convention with $R=1$ and different values of
            $R_{\rm sep}$ to CDF data.}}
 \end{center}
\end{figure}

Our results are shown in fig.~\ref{plot8}.
%
\begin{figure}[p]
 \begin{center}
  {\unitlength1cm
  \begin{picture}(13,8.7)
   \epsfig{file=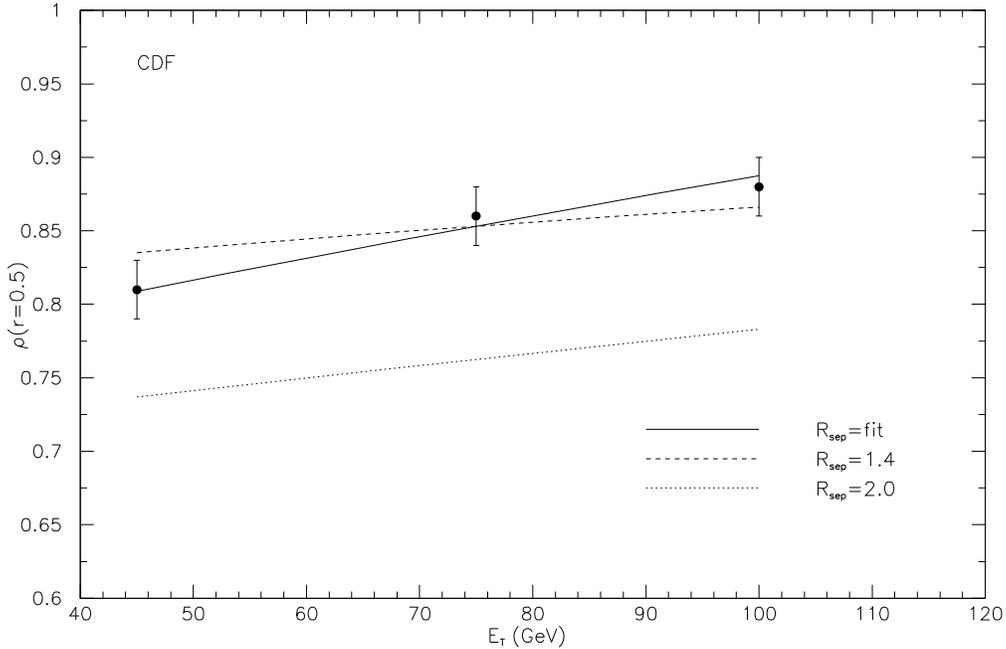,bbllx=520pt,bblly=95pt,bburx=105pt,bbury=710pt,%
           height=13cm,clip=,angle=270}
  \end{picture}}
  \caption{\label{plot9}{\it Jet shape $\rho(E_T)$ for single-jet production in
            $p\overline{p}$ collisions for $r=0.5$ and integrated over
            $0.1\leq|\eta|\leq 0.7$ and the same regions of $E_T$ as
            in the last figure. We compare our results with
            one variable and two fixed values of $R_{\rm sep}$ to CDF data.}}
 \end{center}
\end{figure}
We present three curves, $R_{\rm sep}$ = 1.4 and $R_{\rm sep}$ = 2.0 fixed for
all $E_T$
and a curve with the optimized $R_{\rm sep}$, which is $R_{\rm sep}$ = 1.5,
1.4, and 1.3 for the three $E_T$
bins. Thus $R_{\rm sep}$ increases with decreasing $E_T$
as we expect it from the comparison with the ZEUS data. It is clear
from fig.~\ref{plot8} that $R_{\rm sep}$ = 2.0 describes the data in none of
the $E_T$
bins. Second, a small increase of $R_{\rm sep}$ is needed to explain the data
in the two lower $E_T$ intervals. \\

This is observed also when we plot $\rho (r = 0.5)$ as a function of $E_T$
and compare with the data \cite{X1} in fig.~\ref{plot9}. Although fixed
$R_{\rm sep}$ = 1.4 is in reasonable agreement, the optimized $R_{\rm sep}$
accounts perfectly for the three data points. We remark that the $E_T$
dependence of $\rho$
at $r = 0.5$ in fig.~\ref{plot9} looks very similar to that in fig.~\ref{plot4}
for the ZEUS data. Here also the variation of $R_{\rm sep}$ with $E_T$
was small $(\Delta R_{\rm sep} = 0.1$ similar to
$\Delta R_{\rm sep} = 0.2$ for the CDF data).

\subsection{Comparison with D0 Data}
Last year the D0 collaboration at the TEVATRON presented their
jet shape data \cite{X6} covering a similar kinematic range as
CDF. Their data were used to populate four non-overlapping  jet $E_T$
ranges of 45-70, 70-105, 105-140, and greater than 140 GeV. The jets
were analyzed in a central region
$|\eta| \leq 0.2$
and a forward region
$2.5 \leq |\eta| \leq 3.0$.
In the forward region the data were collected only in the first two
$E_T$
bins. In the experimental analysis jets were reconstructed using a
fixed cone algorithm with $R = 1$. The preclusters were constructed using
the Snowmass jet direction definitions in (1) and (2). After a stable
jet center was found, the jet direction was redefined using the D0
jet direction definition
\begin{equation}
 \eta_J=-\ln(\tan(\theta_J/2)),
\end{equation}
\begin{equation}
 \phi_J=\tan^{-1}(\sum_{i\in{\rm jet}}
 E_{y_i}/\sum_{i\in{\rm jet}}E_{x_i}),
\end{equation}
where the jet polar angle
\begin{equation}
 \theta_J=\tan^{-1}(\sqrt{(\sum_{i\in{\rm jet}}
 E_{x_i})^2+(\sum_{i\in{\rm jet}}E_{y_i})^2}/\sum_{i\in{\rm jet}}E_{z_i}).
\end{equation}
This definition of the jet direction differs from the commonly
used Snowmass definition given in (1) and (2) above. \\

The merging
of overlapping jets was done similarly as in the CDF and ZEUS
analyses. Two jets were merged into one jet if more than 50\% of
the
$E_T$
of the jet with the smaller
$E_T$
was contained in the overlap region. If less than 50\% of the
$E_T$
was contained in the overlap region, the jets were split into
two distinct jets, where the energy in the overlap region was
assigned to the nearest jet. \\

After this procedure the jet
directions were recalculated with the D0 conventions described above.
So the main difference in the D0 cone algorithm is the definition
of the jet direction. This influences both the data and the
theoretical predictions for the jet shape. It was found, however,
that the measured jet shapes hardly change, when instead of the D0
definition for the jet direction the Snowmass definition was used.
In the central region
$|\eta|< 0.2$
the change of
$\rho$
was
$\sim 3\%$
at most (for the inner subcone) and not more than 4\% for jets in
the forward region
$(2.5<|\eta|<3.0)$
\cite{X6}. However, on the theoretical side the jet shapes using
the two jet definitions were much more different. In particular
the forward produced jets are much wider with the D0 definition
than with the Snowmass definition. Furthermore the corresponding
theoretical jet shapes did not agree at all with the experimental
data. Thus, whereas the experimental data are relatively
insensitive to the choice of jet direction definition, the
theoretical predictions vary appreciably. This latter point was
studied further by Glover and Kosower \cite{X19}. They showed
that D0's jet direction definition is not as perturbatively stable
as the Snowmass definition. Therefore they concluded that the D0
recombination scheme should not be applied for the purpose of
making a quantitative comparison with fixed order perturbation
theory. Since the difference of the measured jet shapes between
the two schemes of jet direction definition is so small, we shall
in the following assume that the data of D0 \cite{X6} are
obtained with the Snowmass definition and shall compare them with
our prediction obtained with this algorithm. \\

\begin{figure}[p]
 \begin{center}
  {\unitlength1cm
  \begin{picture}(13,8.7)
   \epsfig{file=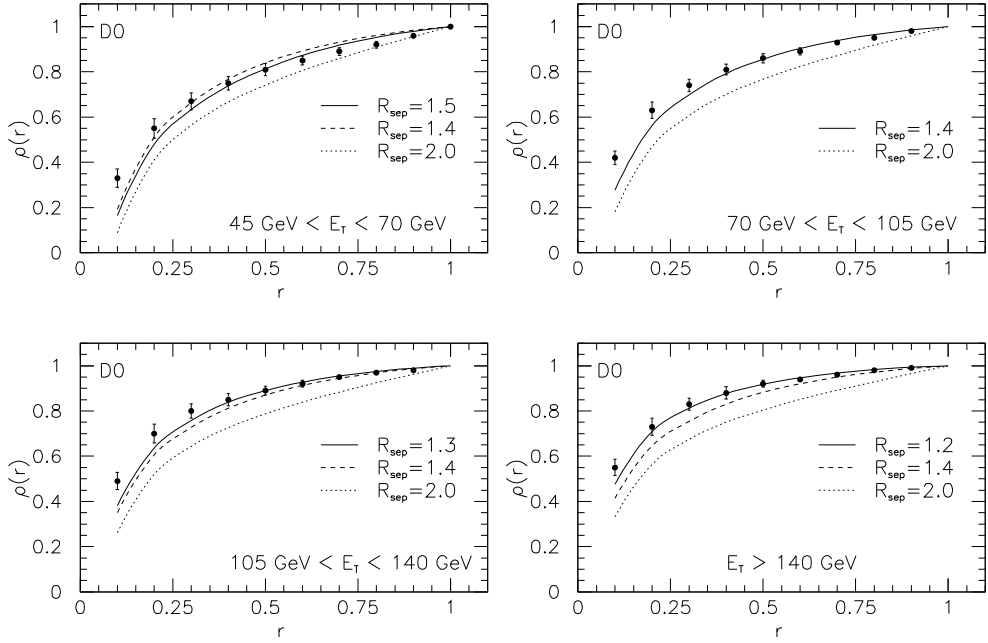,bbllx=520pt,bblly=95pt,bburx=105pt,bbury=710pt,%
           height=13cm,clip=,angle=270}
  \end{picture}}
  \caption{\label{plot10}{\it Jet shape $\rho(r)$ for single-jet production in
            $p\overline{p}$ collisions integrated over $|\eta|<0.2$
            and four different regions of $E_T$. We compare our results using
            the Snowmass convention with $R=1$ and different values of
            $R_{\rm sep}$ to D0 data.}}
 \end{center}
\end{figure}

Our results for $\rho$
as a function of $r$ are plotted in fig.~\ref{plot10},
%
\begin{figure}[p]
 \begin{center}
  {\unitlength1cm
  \begin{picture}(13,5)
   \epsfig{file=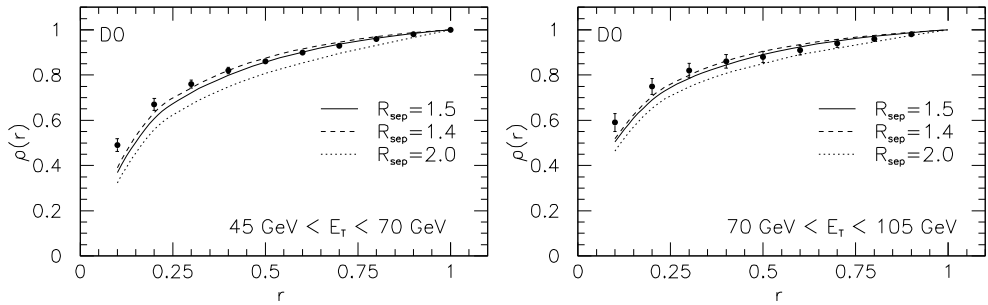,bbllx=300pt,bblly=95pt,bburx=105pt,bbury=710pt,%
           height=13cm,clip=,angle=270}
  \end{picture}}
  \caption{\label{plot11}{\it Jet shape $\rho(r)$ for single-jet production in
            $p\overline{p}$ collisions integrated over $2.5<|\eta|<3.0$
            and the two lower regions of $E_T$. We compare our results using
            the Snowmass convention with $R=1$ and three different values of
            $R_{\rm sep}$ to D0 data.}}
 \end{center}
\end{figure}
separated into the four $E_T$ bins and for $|\eta| < 0.2$.
The curves in fig.~\ref{plot10} are for fixed $R_{\rm sep}$ = 1.4 and
$R_{\rm sep}$ = 2.0 and
for an optimized $R_{\rm sep}$ (full curve) to produce the best fit to the
D0 data. The fitted $R_{\rm sep}$'s vary between 1.5 (lowest $E_T$
bin) to 1.2 for the
$E_T>140$ GeV measurements. So $R_{\rm sep}$ decreases with increasing $E_T$
and $R_{\rm sep}$ = 1.3 for the second largest
bin agrees quite well with the $R_{\rm sep}$ obtained for the CDF data in the
same $E_T$
range. We observe again, that the variation of $R_{\rm sep}$ is small,
$\Delta R_{\rm sep}$
= 0.3 similar to our result obtained from the comparison with the
CDF data. \\

The same plots for jet production in the forward direction,
$2.5 < |\eta| < 3.0$, are shown in fig.~\ref{plot11}. Here we have only two
$E_T$ intervals. The fitted $R_{\rm sep}$ is $R_{\rm sep}$ = 1.5 in both $E_T$
bins, but $R_{\rm sep}$ = 1.4 also gives a reasonably good description. \\

\begin{figure}[ht]
 \begin{center}
  {\unitlength1cm
  \begin{picture}(13,8.7)
   \epsfig{file=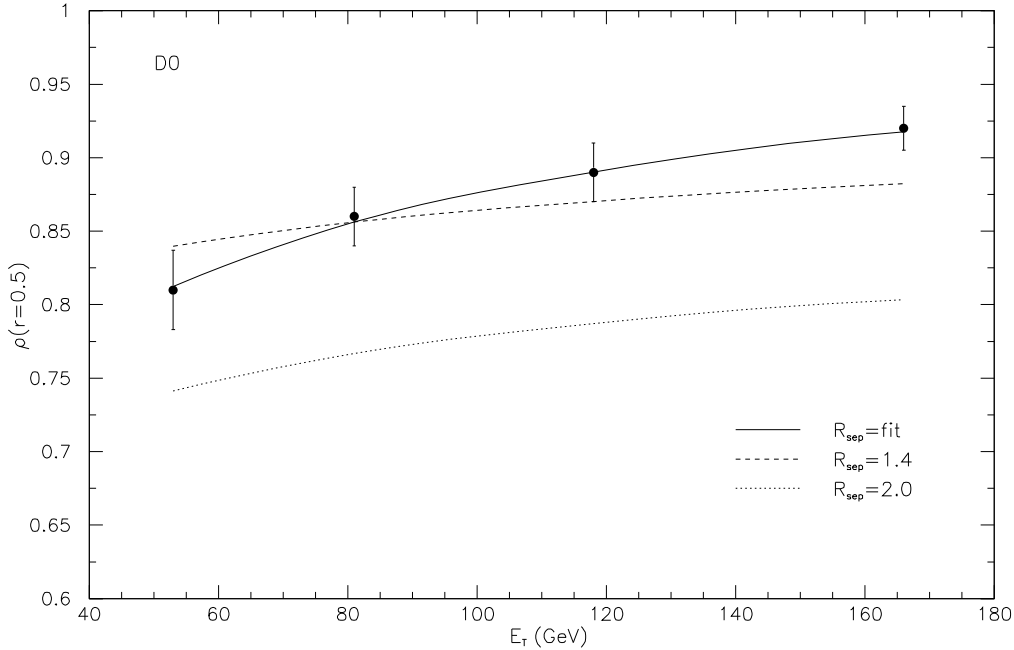,bbllx=520pt,bblly=95pt,bburx=105pt,bbury=710pt,%
           height=13cm,clip=,angle=270}
  \end{picture}}
  \caption{\label{plot12}{\it Jet shape $\rho(E_T)$ for single-jet production
            in $p\overline{p}$ collisions for $r=0.5$ and integrated over
            $|\eta|<0.2$ and the same regions of $E_T$ as
            in fig.~\ref{plot10}. We compare our results with
            one variable and two fixed values of $R_{\rm sep}$ to D0 data.}}
 \end{center}
\end{figure}

If we fix $r$ = 0.5, the corresponding $\rho$ as a function of $E_T$
is seen in fig.~\ref{plot12}
for the case of the central jets. We see that a
slight change of $R_{\rm sep}$ is needed to reproduce the data similar to
the result from fig.~\ref{plot9}. For the D0 data this is even more
convincing than for the CDF data since the range of
$E_T$,
where experimental data exist, is now larger. \\

From this comparison of jet shapes measured in
$p\overline{p}$
collisions at
$\sqrt{s}$
= 1.8 TeV we conclude that we obtain a quantitative description of
the data when we consider a varying parameter $R_{\rm sep}$ which
decreases as a function of $E_T$ and is equal to $R_{\rm sep}$ = 1.3 for
$E_T$ = 100 GeV jets. The variation of $R_{\rm sep}$, needed to explain the
measurements in the considered $E_T$ range, is rather small,
$\Delta R_{\rm sep}\simeq 0.2$
in agreement with the findings for jets produced in $\gamma p$ collision at
$\sqrt{s}$ = 300 GeV. \\

So far we have not mentioned that our theoretical predictions
depend on the choice of the renormalization and factorization
scales which we have set equal to $E_T$. Since the jet profile $1-\rho$
as given in (8) is a LO prediction, i.e. is ${\cal O}(\alpha_s)$,
the dependence on the scales is stronger than in a genuine NLO result
which includes corrections ${\cal O}(\alpha_s^2)$.
According to (8) the quantity $1-\rho$
is a ratio of cross sections. Therefore we expect that the
dependence on the factorization scale cancels to a large extent and $1-\rho$
depends essentially only on the renormalization scale $\mu$. When $\mu$
is decreased, $\alpha_s$ increases, so that $1-\rho$
increases, i.e.~the jets become broader. In principle we could
repeat the analysis by varying
$\mu$
in the interval
$E_T/2 < \mu < 2E_T$
as is usually done. However, such a variation of
$\mu$
has very little influence on
$\rho(r)$
for
$r \geq 0.5$
\cite{X2, X4}. For $r < 0.5$ the change of
$\rho(r)$
is noticeable. Thus the fits to the data might even improve in this
range of $r$, when $R_{\rm sep}$ and
$\mu$
are optimized simultaneously. This would not change the trend of
the $R_{\rm sep}$ variation with decreasing
$E_T$.
Only the optimized $R_{\rm sep}$ values would change slightly. Furthermore,
for small $r < 0.5$, the jet shape
$\rho(r)$
is not predicted very well, since in the region
$r\rightarrow 0$
higher order corrections become more and more important. The change
of
$\rho$
with $R_{\rm sep}$ occurs also for large $r$, where the predictions are more
reliable and the data have smaller errors. From this we conclude
that the $R_{\rm sep}$ values deduced from the data should not be very
sensitive
to the choice of the renormalization scale. The same conclusion can
be drawn concerning our choice of the one-loop formula for
$\alpha_s$
versus the two-loop formula fixing the
$\Lambda$
value through the parton distribution of the proton. Making other
choices is just equivalent to small changes of the renormalization
scale.

\section{Summary}
We have calculated jet shapes in photoproduction superimposing direct and
resolved contributions and in $p\overline{p}$ collisions in next-to-leading
order of perturbation theory. The dependences of the jet shapes on transverse
energy, rapidity, and inner cone extension were compared with recent data from
ZEUS at HERA and from CDF and D0 at the TEVATRON. The numerical results were
obtained using the Snowmass cone algorithm yielding good agreement with the
data, if the problem of merging overlapping jets and additional jet broadening
effects are taken into account by
varying the parameter $R_{\rm sep}$ as a function of transverse energy
and rapidity. The optimal values and variations of $R_{\rm sep}$ with $E_T$
were found to be rather similar in low $Q^2$ $ep$ and in $p\overline{p}$
collisions. Furthermore, it was shown that the convex shape of the jet profile
rapidity distribution in photoproduction is due to the direct photon
contribution.

\section*{Acknowledgements}
We would like to thank C. Glasman and J. Terron for making the ZEUS data
available to us prior to publication.

\end{document}